# Gaia Shows That Messier 40 is Definitely Not a Binary Star

*Michael R. Merrifield, Meghan E. Gray and Brady Haran*
*School of Physics & Astronomy*
*University of Nottingham*
*Nottingham*
*NG7 2RD*

Messier 40 has always been something of an oddity in the Messier Catalogue, since it is just a pair of stars rather than a nebulous object.[1]  It seems to have ended up in the list because Charles Messier was following up on an observation by Hevelius[2] of a nebulous binary star in this vicinity.  Although Messier identified no nebulosity (and, in fact, had found a different pair of stars to those seen by Hevelius), he added the pairing to his catalogue for completeness.  Surprisingly, the identification was then lost until 1966, when the pair of stars were noted as almost certainly HD 238107 and HD 238108.[3]

Little was known about this pair of stars, but a thorough analysis in 2002 noted that the proper motions derived from the Tycho-2 Catalogue and earlier observations indicated that these stars were likely unrelated objects passing each other on the sky.[4]  Support for this interpretation came from the spectroscopic parallaxes estimated for these objects of 590±230 pc and 170±70 pc respectively.[4]  However, the error bars were uncomfortably large, and the distances sufficiently great that Hipparcos data placed no significant constraint on their astrometric parallax distances, so the possibility still remained that the stars could form a wide physical binary.[4]

The first data release from the Gaia satellite[5] definitively puts paid to this interpretation.  The combined Hipparcos/Gaia astrometric solution gives parallaxes of 2.87±0.24 milliarcseconds and 7.13±0.24 milliarcseconds, translating to distance estimates of 350±30 pc and 140±5 pc for the two objects.  Messier 40 comprises a pair of entirely unrelated stars.